\documentclass[journal]{IEEEtran}
\ifCLASSINFOpdf
  \usepackage[pdftex]{graphicx}
  \graphicspath{../figtex/}
  \DeclareGraphicsExtensions{.jpeg,.png}
\else
\fi
\usepackage{amssymb,amsmath,amsthm,amsfonts}
\DeclareMathAlphabet\mathbfcal{OMS}{cmsy}{b}{n}
\usepackage{algorithm,algorithmic}

\usepackage{array}

\ifCLASSOPTIONcompsoc
  \usepackage[caption=false,font=normalsize,labelfont=sf,textfont=sf]{subfig}
\else
  \usepackage[caption=false,font=footnotesize]{subfig}
\fi
\begin{document}
%

\title{Source Localization of an Unknown Transmission in Dense Multipath Environments}

%
%
%

\author{Asaf~Afriat,
        Dan~Raphaeli,
        Oded~Bialer
}

\maketitle

\begin{abstract}
    Accurately estimating the position of a wireless emitter in a multipath environment based on samples received at various base stations (in known locations) has been extensively explored in the literature. Existing approaches often assume that the emitted signal is known to the location system, while in some applications, such as locating surveillance or intelligence systems, it usually remains unknown. In this paper, we propose a novel estimator for determining the position of an emitter transmitting an unknown signal in a dense multipath environment with a given power-delay profile. We also derive the Cramer–Rao lower bound (CRLB) to evaluate the estimator's performance. Our approach is based on approximating the dense multipath channel in the frequency domain as a Gaussian random vector using the central limit theorem, formulating a log-likelihood cost function for the position and some features of the transmitted signal, and applying a maximum search over both. The optimization problem is non-convex and has no known analytical solutions, which makes it computationally infeasible for multidimensional brute-force search. To address this challenge, we developed a practical optimization algorithm that overcomes the computational complexity, using reasonable approximations, that provides a feasible position estimator. Through extensive evaluations, we demonstrate that the proposed estimator outperforms other state-of-the-art estimators. Moreover, as the number of base stations and SNR increase, our estimator approaches the CRLB, indicating its effectiveness and efficiency.
\end{abstract}


%
\IEEEpeerreviewmaketitle

\section{Introduction} \label{sec:intro}
%
%
%
%

\IEEEPARstart{A}{ccurately} estimating the location of a wireless transmitting device holds great significance in various applications, including navigation, rescue missions, traffic management, inventory tracking, patient monitoring, and more. To this day, many positioning methods are based on the relation between the device position and the time-of-arrival (TOA) of its transmission at multiple receiving base stations (BSs) or the time-difference-of-arrival (TDOA) between them \cite{intro:zekavat2021overview,intro:gezici2005localization}.

In a free-space environment, the transmitted signal follows a single path to each BS, simplifying the measurement of TOA or TDOA. However, in congested environments such as indoor offices, urban streets, forests, etc., the signal undergoes reflections from numerous surrounding objects, causing it to propagate through multiple paths to each BS. Consequently, the received signal becomes a combination of multiple replicas of the transmitted signal, each with varying delays and attenuations. As a result, accurately estimating the TOA or TDOA between BSs becomes exceptionally challenging.

The subject of modeling multipath channels has been extensively researched, leading to comprehensive studies and baseband models \cite{intro:saleh1987statistical,intro:molisch2006comprehensive}. These models represent multipath channels as clusters of statistically independent complex coefficients with variances that decay exponentially. The coefficients follow Rayleigh or Nakagami distributions, while their arrival times are modeled as Poisson distributions.

The literature contains various methods for estimating the position of a wireless device using TOA or TDOA measurements in a multipath environment \cite{intro:qi2006time,intro:lymberopoulos2015realistic}. However, many of these approaches assume prior knowledge of the transmitted signal by the estimator. Yet, certain applications require estimating the position of a wireless device without knowledge of the transmitted signal. For instance, surveillance or intelligence systems often use unique communication protocols that are unknown to the location system. Additionally, many communication systems transmit signals with a known preamble followed by unknown data symbols, where the preamble duration is significantly shorter compared to the data duration. In such cases, relying solely on the short preamble for position estimation might yield a low signal-to-noise ratio (SNR), making accurate position estimation challenging. However, leveraging the unknown part of the transmission for position estimation could yield a substantial SNR improvement, enhancing the accuracy of the results.

Tirer and Weiss \cite{MUSIC:tirer2016high} derived the maximum likelihood (ML) position estimator for scenarios involving an unknown transmit signal in a line-of-sight (LOS) channel with a single path. This ML estimator (MLE) can also be applied in multipath channels, provided that the LOS path is considerably stronger than the non-LOS (NLOS) paths or when the multipath delay spread is significantly shorter than the inverse of the signal bandwidth. However, in environments like indoor spaces or urban areas, the channel often exhibits a relatively large delay spread, and the intensity of NLOS paths can be significant compared to that of the LOS path. In such environments, utilizing an MLE derived for LOS channels may lead to poor performance. 

Jianping \textit{et al.} \cite{MUSIC:du2018music} devised position estimation techniques for a wireless system consisting of BSs and transponders in known locations, where the source to be located is transmitting an unknown signal. In this particular setup, the multipath channel is man-made as it is assumed that the transmission propagates in a direct path from the source to each transponder and from each transponder to each BS, without any other propagation paths present. Their study introduces two estimation approaches: a Multiple Signal Classification (MUSIC)-based method and an ML approach. In this paper, we offer a way to generalize the MUSIC-based estimator to work in an unknown environment (See Section \ref{sec:results}), while the ML estimator seems to be constrained to a man-made multipath channel only.

Another technique described in the literature \cite{kikuchi2006blind,n2018enhancement} for locating the source of an unknown transmission in a multipath environment involves employing ray-tracing analysis, a method used to calculate the propagation paths of rays within a specific environment. Though this approach has been shown to achieve accurate positioning results, it relies on pre-acquired accurate three-dimensional terrain data for each tested environment.

More recently, Kehui \textit{et al.} \cite{zhu2022cross} proposed a position estimator that eliminates the need for pre-collected data. This estimator is designed to operate in an unknown multipath environment with an unknown transmit signal. The estimator treats one of the BSs as a 'reference' and calculates the measured cross-spectrum between the received signal in the reference BS and the signals received at other BS. By employing a MUSIC-based approach over the cross-spectrum measurements, the estimator achieves superior performance compared to reference methods. It is important to note that in the mathematical derivation of this estimator, the reference signal is assumed to be propagating through a single LOS path. However, they have empirically shown that even when the reference BS encounters a small number of NLOS arrivals, the estimator still demonstrates excellent performance. Yet, in densely populated environments with many propagation paths, the LOS approximation about the reference BS may not hold, leading to severe performance degradation, as shown in Section \ref{sec:results} using extensive simulations.

In this paper, we present a novel estimator for determining the position of a wireless device that transmits an unknown signal in a dense multipath environment, given its power-delay profile (PDP). We start by approximating the channel in the frequency domain to be a Gaussian random vector. Then we can formulate a log-likelihood cost function for both the position and some features of the transmitted signal. To obtain the position we need to maximize the cost function over both. This optimization problem is non-convex and has no known analytical solution. Hence, we developed a practical solution for this problem by first estimating the magnitudes of the transmitted signal independently from its phases and the source position, then converting the problem into a familiar non-convex optimization problem that has been discussed in the literature \cite{liu2017gpm,waldspurger2015phase}. This optimization is repeated for each position hypothesis and the optimized position is found.

Furthermore, we developed a method of complexity reduction in the case of long observation periods and derived the Cramér–Rao lower bound (CRLB) for the estimation problem at hand.

The proposed estimator significantly outperforms other mentioned reference estimators in various settings, and approaches the CRLB when the number of BSs is large and the SNR is high.



\section{System Model and Problem Formulation} \label{sec:sysmod}
Consider a single transmitter positioned at an unknown location in three-dimensional space, denoted as $\boldsymbol{q}\in\mathbb{R}^3$. Additionally, there are $M$ BSs located at known positions, represented as $\boldsymbol{p}_m\in\mathbb{R}^3$, where the index $m$ ranges from $0$ to $M-1$. Both the transmitter and BSs remain stationary, and all BSs are assumed to be time synchronized. Let us use a base-band signal representation in our model. The channel between the transmitter to the $m$th BS is considered to be a time-invariant multipath channel and is represented as
\begin{equation}\label{eq:sysmodel:ch_time}
    h_m\left(t\right)=\sum_{l=0}^{L_m-1}{\alpha_{m,l}\delta\left(t-\tau_{m,l}\right)},
\end{equation}
where $\delta\left(t\right)$ is the Dirac delta function, $L_m$ is the number of multipath arrivals in the $m$th channel, $\alpha_{m,l}$ is the complex gain of the $l$th arrival to the $m$th BS and $\tau_{m,l}$ is its delay. Without loss of generality, we index the arrivals in increasing order of delay. We assume that the first arrival ($l=0$) comes from the LOS path and that $\alpha_{m,l}$ and $\tau_{m,l}$ are statistically independent random variables, both in $m$ and $l$ indices. Namely, the multipath components per channel are statistically independent and so are any two channels (to different BSs). The probability distribution of the channel is unknown, however, its PDP can be measured or modeled (see Section \ref{sec:mle:deter}) and is considered known. 

The complex base-band signal, observed by the $m$th BS, in time interval $t\in\left[0,T_{obs}\right]$, is therefore
\begin{equation}\label{eq:sysmodel:sig_time}
    r_m\left(t\right)=\sum_{l=0}^{L_m-1}{\alpha_{m,l}s\left(t-\tau_{m,l}\right)}+n_m\left(t\right),
\end{equation}
where $s\left(t\right)$ is the transmitted signal, and $n_m\left(t\right)$ is an additive white Gaussian noise (AWGN). Let $s(t)$ have bandwidth $W$. At each BS the observed signal, $r_m\left(t\right)$, is passed through a pre-sampling low-pass filter (LPF) and sampled at rate $F_s=T_s^{-1}$, satisfying the condition $F_s \ge W$. The number of received samples is denoted by $N_s$, which gives an observation time of $T_{obs}=N_sT_s$. The complete observation interval can be divided into $D$ contiguous observation 'windows' of length $T_w=T_{obs}/D$. The number of samples within each window is indicated by $K$ ($N_s=DK$).

\begin{figure}
    \centering
    \includegraphics[scale=0.32]{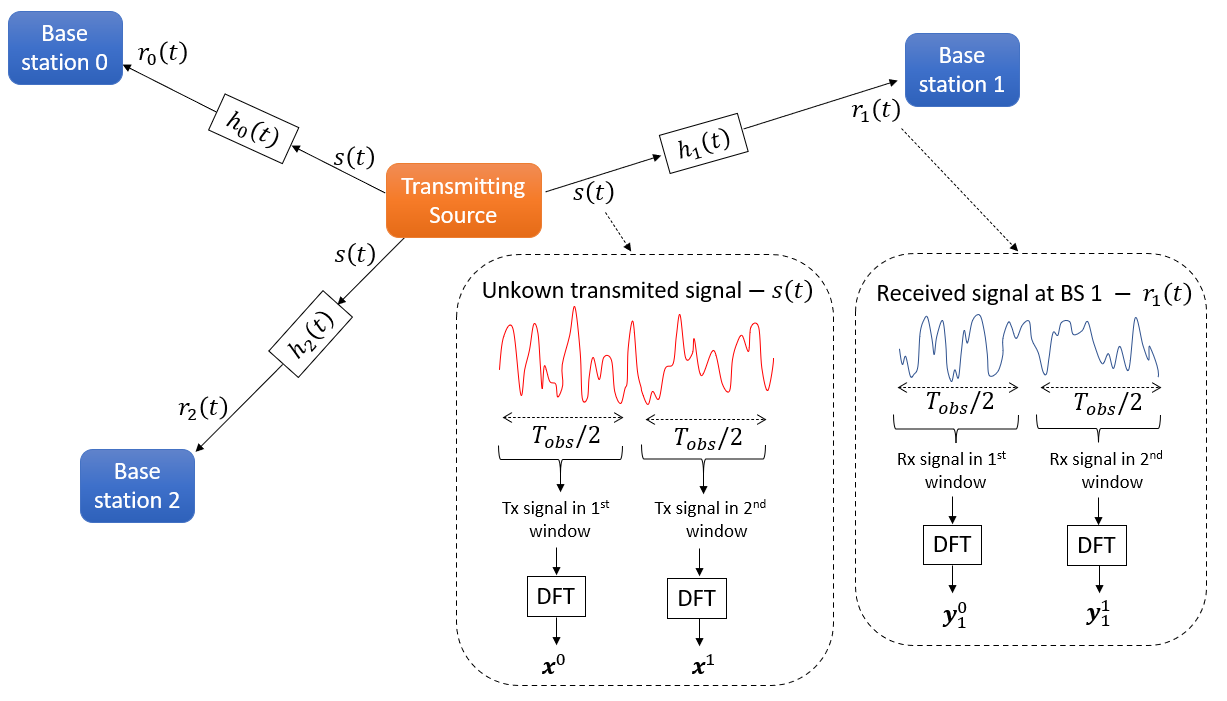}
    \caption{Base-band system model for three BSs. The source transmits an unknown signal, $s(t)$, that propagates through a different channel to each BS ($h_{0,1,2}$). In this illustration, the entire observation interval, $T_{obs}$, is partitioned into two windows. The DFTs of the transmitted signal in the two windows are denoted by $\boldsymbol{x}^{0,1}$. The DFTs of the signal received in the two windows, at the base station indexed as $m=1$, are denoted by $\boldsymbol{y}_1^0$ and $\boldsymbol{y}_1^1$.}
    \label{fig:sysmod:baseband}
\end{figure}

Let $\boldsymbol{y}_m^d\in\mathbb{C}^K$ be the discrete Fourier transform (DFT) of the sampled signal at the $d$th window, as illustrated in Fig. \ref{fig:sysmod:baseband}. It is assumed that all BSs initiate and terminate the observation windows at the same time. $T_w$ is set to be much larger than the channel's delay spread so that the edge effects in the analysis of the DFT are negligible. Under these conditions, the $k$th DFT coefficient of $\boldsymbol{y}_m^d$ is given by
\begin{equation}\label{eq:sysmodel:sig_freq}
    y_m^{d}[k]=\sum_{l=0}^{L_m-1}{\alpha_{m,l}e^{-j2\pi f_k\tau_{m,l}}\cdot x^{d}[k]}+v_m^{d}[k],
\end{equation}
where $x^{d}[k]$ and $v_m^{d}[k]$ are the $k$th DFT coefficients of $s(t)$ and $n_m(t)$ at the $d$th window, respectively, and $f_k\triangleq kF_s/K$. 

Throughout this paper, we shall use the superscripts $(\cdot)^\ast$, $(\cdot)^T$, and $(\cdot)^\dag$ to represent the element-wise conjugate, matrix transpose, and matrix conjugate-transpose operators, respectively. We express (\ref{eq:sysmodel:sig_freq}) in matrix notation as follows
\begin{equation} \label{eq:sysmodel:sig_freq_mat}
    \boldsymbol{y}_m=\boldsymbol{X}\sum_{l=0}^{L_m-1}{\alpha_{m,l}\boldsymbol{g}_{\tau_{m,l}}}+\boldsymbol{v}_m,
\end{equation}
where
\begin{equation}\label{eq:sysmodel:g_vec_def}
    \boldsymbol{g}_\tau\triangleq\begin{bmatrix}e^{-j2\pi f_0\tau},&\ldots,&e^{-j2\pi f_{K-1}\tau}\end{bmatrix}^T,
\end{equation}
\begin{gather*}
    \boldsymbol{y}_m\triangleq\begin{bmatrix}{\boldsymbol{y}_m^{0}}^T,&\ldots,&{\boldsymbol{y}_m^{D-1}}^T\end{bmatrix}^T, \\
    \boldsymbol{y}_m^{d}\triangleq\begin{bmatrix}y_m^{d}[0],&\ldots,&y_m^{d}[K-1]\end{bmatrix}^T, \\
    \boldsymbol{v}_m\triangleq\begin{bmatrix}{\boldsymbol{v}_m^{0}}^T,&\ldots,&{\boldsymbol{v}_m^{D-1}}^T\end{bmatrix}^T, \\
    \boldsymbol{v}_m^{d}\triangleq\begin{bmatrix}v_m^{d}[0],&\ldots,&v_m^{d}[K-1]\end{bmatrix}^T, \\
    \boldsymbol{X}\triangleq\left[\boldsymbol{X}^0,\ldots,\boldsymbol{X}^{D-1}\right]^T, \\
    \boldsymbol{X}^{d}\triangleq diag\left\{\boldsymbol{x}^{d}\right\}=\begin{bmatrix}x^{d}[0]&\cdots&0\\\vdots&\ddots&\vdots\\0&\cdots&x^{d}[K-1]\end{bmatrix}, \\
    \boldsymbol{x}^{d}\triangleq\begin{bmatrix}x^{d}[0],&\ldots,&x^{d}[K-1]\end{bmatrix}^T.
\end{gather*} 

Let $\boldsymbol{x}\triangleq\begin{bmatrix}{\boldsymbol{x}^{0}}^T,&\ldots,&{\boldsymbol{x}^{D-1}}^T\end{bmatrix}^T$ be a vector containing the DFT coefficients of transmitted signals from all $D$ observation windows. We consider a system in which the transmitted signal, $\boldsymbol{x}$, is unknown and the multipath is dense, i.e., the number of multipath arrivals, $L_m$, within $W^{-1}$ is large. The first arrival time, $\tau_{m,0}$, comes from the LOS path and is a function of the distance between the transmitter and the BS, expressed by
\begin{equation}\label{eq:mle:toa}
    \tau_{m,0}(\boldsymbol{q})=\frac{\|\boldsymbol{q}-\boldsymbol{p}_m\|}{c},
\end{equation}
where $\|\cdot\|$ denotes the Euclidean norm, and $c$ is the speed of light. Therefore, the received vector, $\boldsymbol{y}_m$, is a function of the transmitter's position.

The problem at hand is estimating the transmitter's location, $\boldsymbol{q}$, given the observations from all BSs, $\left\{\boldsymbol{y}_m\right\}^{M-1}_{m=0}$, and their positions, $\left\{\boldsymbol{p}_m\right\}^{M-1}_{m=0}$, when the transmitted signal, $\boldsymbol{x}$, and the multipath channel components, $\alpha_{m,l}$, $\tau_{m,l}$ and $L_m$, are unknown.

\section{Position Estimator Derivation} \label{sec:mle}
Looking back at (\ref{eq:sysmodel:sig_freq_mat}) we can rewrite it as
\begin{equation} \label{eq:mle:sig_freq_mat}
    \boldsymbol{y}_m=\boldsymbol{X}\boldsymbol{G}_m(\boldsymbol{q})\cdot\boldsymbol{\eta}_m+\boldsymbol{v}_m,
\end{equation}
where
\begin{equation} \label{eq:mle:ch_vec}
    \boldsymbol{\eta}_m\triangleq\sum_{l=0}^{L_m-1}{\alpha_{m,l}\boldsymbol{g}_{\tau'_{m,l}}},
\end{equation}
\begin{gather*}
    \tau'_{m,l}\triangleq\tau_{m,l}-\tau_{m,0}, \\  \boldsymbol{G}_m(\boldsymbol{q})\triangleq diag\left\{\boldsymbol{g}_{\tau_{m,0}(\boldsymbol{q})}\right\}.
\end{gather*}
Note that $\tau'_{m,l}\geq0$ is the time difference between arrivals of the $l$th and LOS paths. It is governed by the position of reflecting objects in the environment which are scattered randomly such that $\tau'_{m,l}$ is assumed to be independent of the transmitter’s position. Therefore, $\boldsymbol{g}_{\tau'_{m,l}}$ is considered statistically independent of $\boldsymbol{q}$, $\boldsymbol{p}_m$ and $\tau_{m,0}$.

The paths' gains, $\alpha_{m,l}$, and relative arrival times, $\tau'_{{m,l}}$, are statistically independent, which means that $\boldsymbol{\eta}_m$ is a sum of $L_m$ statistically independent random vectors. Let $E\{\cdot\}$ denote the mean-value operator. We assume $E\{\alpha_{m,l}\}=0$, and so $E\left\{\boldsymbol{\eta}_m\right\}=\sum_{l=0}^{L_m-1}{E\left\{\alpha_{m,l}\right\} E\left\{\boldsymbol{g}_{{\tau'}_{m,l}}\right\}=\boldsymbol{0}}$. The central limit theorem (CLT) \cite{BOOK:mittelhammer2013mathematical} states that the distribution of a sum of independent random vectors converges to the distribution of a Gaussian vector, for a large enough set of vectors. Therefore, in dense multipath environments, where the average number of arrivals within $W^{-1}$ is high \cite{bialer2012Efficient}, we use the CLT to approximate $\boldsymbol{\eta}_m$, and in turn $\boldsymbol{y}_m$, to be complex Gaussian random vectors given $\boldsymbol{x}$ and $\boldsymbol{q}$.

According to (\ref{eq:mle:sig_freq_mat}), $E\{\boldsymbol{y}_m|\boldsymbol{q},\boldsymbol{x}\}=\boldsymbol{0}$ is implied form the fact that $E\{\boldsymbol{\eta_m}\}=\boldsymbol{0}$ and $E\{\boldsymbol{v_m}\}=\boldsymbol{0}$. By making use of the Gaussian approximation, the probability density function (PDF) of $\boldsymbol{y}_m$ given $\boldsymbol{q}$ and $\boldsymbol{x}$ is
\begin{equation} \label{eq:mle:pdf}
    \mathbf{f}\left(\boldsymbol{y}_m\middle|\boldsymbol{q},\boldsymbol{x}\right)=\frac{1}{\pi^{KD}\left|\boldsymbol{R}_m\right|}\cdot\exp\left\{-\boldsymbol{y}_m^\dag\boldsymbol{R}_m^{-1}\boldsymbol{y}_m\right\},
\end{equation}
where 
\begin{equation}\label{eq:mle:cov}
    \boldsymbol{R}_m(\boldsymbol{q})\triangleq \boldsymbol{X}\boldsymbol{G}_m(\boldsymbol{q})\boldsymbol{H}_m{\boldsymbol{G}_m}^\dag(\boldsymbol{q})\boldsymbol{X}^\dag+\sigma_v^2\boldsymbol{I},
\end{equation}
\begin{equation*}
    \boldsymbol{H}_m \triangleq E\left\{\boldsymbol{\eta}_m\boldsymbol{\eta}_m^\dag\right\},
\end{equation*}
with $\mathbf{f}(\cdot)$ denotes a PDF, $\boldsymbol{I}$ is the identity matrix and $\sigma_v^2$ is the variance of the AWGN. Note that $\boldsymbol{H}_m$ is the channel covariance matrix corrected for the LOS delay, i.e. setting $\tau_{m,0}=0$, hence, it is positive semi-definite and can be decomposed as follows
\begin{equation}\label{eq:mle:ch_cov}
    \boldsymbol{H}_m=\boldsymbol{U}_m\boldsymbol{U}_m^\dag.
\end{equation}
By substituting (\ref{eq:mle:ch_cov}) into (\ref{eq:mle:cov}) we get
\begin{equation}\label{eq:mle:cov2}
\boldsymbol{R}_m=\boldsymbol{X}\boldsymbol{G}_m\boldsymbol{U}_m\boldsymbol{U}_m^\dag\boldsymbol{G}_m^\dag\boldsymbol{X}^\dag+\sigma_v^2\boldsymbol{I}.
\end{equation}
Making use of the Woodbury matrix identity and the matrix determinant lemma, we obtain
\begin{equation}
    \begin{split}
        \boldsymbol{R}_m^{-1}=\sigma_v^{-2}\boldsymbol{I}-\sigma_v^{-4}\boldsymbol{X}\boldsymbol{G}_m\boldsymbol{U}_m[\boldsymbol{I}+\ldots \\
        +\sigma_v^{-2}\boldsymbol{U}_m^\dag\boldsymbol{X}^\dag\boldsymbol{X}\boldsymbol{U}_m]^{-1}\boldsymbol{U}_m^\dag\boldsymbol{G}_m^\dag\boldsymbol{X}^\dag,
    \end{split} \label{eq:mle:cov_inv}
\end{equation}
and
\begin{equation}
    \left|\boldsymbol{R}_m\right|=\left|\boldsymbol{I}+\sigma_v^{-2}\boldsymbol{U}_m^\dag\boldsymbol{X}^\dag\boldsymbol{X}\boldsymbol{U}_m\right|\cdot\left|\sigma_v^2\boldsymbol{I}\right|. \label{eq:mle:cov_det}
\end{equation}

Since different channels are statistically independent, the received signals from different BSs, $\boldsymbol{y}_m$, are also statistically independent, given $\boldsymbol{x}$ and $\boldsymbol{q}$. Thus using (\ref{eq:mle:pdf}), we can express the joint PDF of received signals from all BSs as
\begin{equation}\label{eq:mle:pdf_tot}
\begin{split}
    &\mathbf{f}\left(\boldsymbol{y}_0,..,\boldsymbol{y}_{M-1}\middle|\boldsymbol{q},\boldsymbol{x}\right)= \\ &=\left(\prod_{m=0}^{M-1}{\pi^{KD}\left|\boldsymbol{R}_m\right|}\right)^{-1}\exp\left\{-\sum_{m=0}^{M-1}{\boldsymbol{y}_m^\dag\boldsymbol{R}_m^{-1}\boldsymbol{y}_m}\right\}.
\end{split}
\end{equation}
Substituting (\ref{eq:mle:cov_inv}-\ref{eq:mle:cov_det}) into (\ref{eq:mle:pdf_tot}), applying $\ln\left\{\cdot\right\}$ to the result and discarding constant terms, we obtain the cost function for position estimation under the Gaussian model approximation for an unknown signal in a multipath environment, given by
\begin{equation*}
    {\hat{\boldsymbol{q}}}=\arg{\max\limits_{\boldsymbol{q}}\ {C_0(\boldsymbol{q})}},
\end{equation*} 
\begin{equation} \label{eq:mle:usge_cf}
    C_0(\boldsymbol{q})\triangleq\max\limits_{\boldsymbol{x}}{\sum_{m=0}^{M-1}{a_m\left(\boldsymbol{y}_m,\boldsymbol{x},\boldsymbol{q}\right)+b_m\left(\boldsymbol{x}\right)}}, 
\end{equation}
where
\begin{gather}
    \begin{split}
        a_m\left(\boldsymbol{y}_m,\boldsymbol{x},\boldsymbol{q}\right)\triangleq\sigma_v^{-4}\boldsymbol{y}_m^\dag\boldsymbol{X}\boldsymbol{G}_m(\boldsymbol{q})\boldsymbol{U}_m\Bigl(\boldsymbol{I}+\ldots\\
        +\sigma_v^{-2}\boldsymbol{U}_m^\dag\boldsymbol{X}^\dag\boldsymbol{X}\boldsymbol{U}_m\Bigr)^{-1}\boldsymbol{U}_m^\dag\boldsymbol{G}_m^\dag(\boldsymbol{q})\boldsymbol{X}^\dag\boldsymbol{y}_m,
    \end{split} \label{eq:mle:usge_a}\\
    b_m\left(\boldsymbol{x}\right)\triangleq-\ln{\left(\left|\boldsymbol{I}+\sigma_v^{-2}\boldsymbol{U}_m^\dag\boldsymbol{X}^\dag\boldsymbol{X}\boldsymbol{U}_m\right|\right)}. \label{eq:mle:usge_b}
\end{gather}

Unfortunately, (\ref{eq:mle:usge_cf}) has no known analytical solution. Moreover, the cost function in (\ref{eq:mle:usge_cf}) is non-convex, hence applying convex optimization solutions is prone to reach local maxima near the initial estimation point. Alternatively, solving (\ref{eq:mle:usge_cf}) with a naive grid search over the unknown parameters, which are the transmitter position, $\boldsymbol{q}$, and transmitted signal, $\boldsymbol{x}$, requires a multidimensional search over a large dimension and hence is unfeasible. In Section \ref{sec:usage}, we resolve this issue by developing a practical algorithm for solving an approximation of (\ref{eq:mle:usge_cf}).

\subsection{Determining $\boldsymbol{H}_m$} \label{sec:mle:deter}
Our proposed estimator requires knowing $\boldsymbol{H}_m$, the covariance matrix of the channel in the frequency domain. One way to obtain it is by empirical measurements done in advance, collecting the received signal of a known transmission from numerous source locations with known TOAs, however, this could be a complex task to perform. Instead, we use a method suggested in \cite{bialer2017Robust} which relies on the channel's PDP.

According to this method, we consider the multipath channel to consist of $N_h$ arrivals confined to a finite time grid with spacing $\Delta \tau$, where $\Delta\tau\ll W^{-1}$ and $(N_h-1)\Delta\tau$ is larger than the maximal delay spread. Then $\boldsymbol{H}_m$ can be expressed as
\begin{equation} \label{eq:mle:ch_cov_grid}
    \boldsymbol{H}_m=\mathbfcal{G}\boldsymbol{\Lambda}_m\mathbfcal{G}^\dag,
\end{equation}
where
\begin{gather*}
    \mathbfcal{G}\triangleq\begin{bmatrix}\boldsymbol{g}_{0},&\boldsymbol{g}_{\Delta\tau},\ldots,&\boldsymbol{g}_{(N_h-1)\Delta\tau}\end{bmatrix},\\
    \boldsymbol{\Lambda}_{m}\triangleq diag\left\{\sigma_{m,0}^2,\ldots,\sigma_{m,N_h-1}^2\right\},\\
    \sigma_{m,n}^2\triangleq E\left\{\left|h_m\left(n\cdot\Delta\tau\right)\right|^2\ |\tau_{m,0}=0\right\}.
\end{gather*}
Note that $\sigma_{m,n}^2$ represents the channel PDP when $\tau_{m,0}=0$.

The PDP itself can be extracted in various methods, such as: 1) empirical measurements at the desired environment, usually using transmissions of narrow pulses from known locations \cite{bialer2012Efficient}; 2) using known channel models like the ultra-wide-band (UWB) IEEE 802.15.4a channel models \cite{uwbchan1,uwbchan2,uwbchan3}; or 3) using a simplified PDP model (e.g., exponential decay) that is governed by a small number of parameters which can be pre-determined by known environment characteristics, as shown in \cite{bialer2017Robust}. Lastly, to use the decomposed form given in (\ref{eq:mle:ch_cov}) we simply set $\boldsymbol{U}_m=\mathbfcal{G}\boldsymbol{\Lambda}_m^{1/2}$.

\section{Practical Optimization Algorithm} \label{sec:usage}
In this section, we develop an efficient algorithm for finding an approximate solution for the maximization problem (\ref{eq:mle:usge_cf}). We start by developing an estimator for the magnitudes of the transmitted signal DFT samples, $|x^{d}[k]|$, independent of their phases and the transmitter position. Then, given the estimated magnitudes, we turn (\ref{eq:mle:usge_cf}) into maximization over the transmit signal phases, $\angle x^{d}[k]$, and solve this problem with an efficient optimization algorithm. In Section \ref{sec:usage:step1} we present the transmit signal magnitudes estimator and in Section \ref{sec:usage:step2} the position and transmit signal phases estimator. Furthermore, in Section \ref{sec:usage:optcomp} we suggest a complexity reduction method based on coherently combining the samples from different observation windows.

It is worth noting that in some practical cases, the magnitudes of the signal's DFT are known, while the phases are not. For example, in orthogonal frequency division multiplexing (OFDM) transmission with phase shift keying (PSK) modulation the data is transmitted in frequency bins that have constant magnitude, due to the PSK modulation. However, the observation windows must match the OFDM frames in all their parameters.

\subsection{Transmit Signal Magnitudes Estimation} \label{sec:usage:step1}
The transmit signal vector, $\boldsymbol{x}$, can be decomposed to magnitude and phase components as follows
\begin{equation}\label{eq:usage:step1:sig_splitmag}
    \boldsymbol{x}=\boldsymbol{\Gamma}\cdot\boldsymbol{\gamma},
\end{equation}
where
\begin{gather*}
    \boldsymbol{\Gamma}\triangleq diag\left\{\left|\boldsymbol{x}\right|\right\}, \\
    \boldsymbol{\gamma} \triangleq \exp\left\{j\angle \boldsymbol{x}\right\},
\end{gather*}
with $\angle\{\cdot\}$ and $\exp\{\cdot\}$ representing the element-wise complex phase and exponential operators, respectively, and with $|\cdot|$ representing the element-wise absolute value when applied over a vector (not to be confused with the matrix determinant operation). We estimate the magnitude of the $k$th DFT coefficient of the transmitted signal at the $d$th window by averaging the energy of the received signals from all BSs, expressed by
\begin{equation} \label{eq:usage:step1:magest}
    \hat{\Gamma}\left[k+K d,k+K d\right]=\left|\hat{x}^{d}[k]\right| = \sqrt{\frac{1}{M}\sum_{m=0}^{M-1}\frac{{\left|{y}_m^{d}[k]\right|}^2}{H_m[k,k]}},
\end{equation}
where $\hat{\boldsymbol{x}}$ represents the estimation of $\boldsymbol{x}$. Additionally, $\hat{\boldsymbol{\Gamma}}$ and $\hat{\boldsymbol{\gamma}}$ represent the terms of magnitudes and phases of the estimated $\hat{\boldsymbol{x}}$, respectively, corresponding to $\boldsymbol{\Gamma}$ and $\boldsymbol{\gamma}$.

We note that more advanced magnitude estimators were also considered, however, they showed no substantial performance advantage in our empirical position estimation tests.

\subsection{Transmit Signal Phases and Position Estimation} \label{sec:usage:step2}
By substituting the estimated transmit signal magnitudes in (\ref{eq:usage:step1:magest}) into (\ref{eq:usage:step1:sig_splitmag}), then substituting the result into (\ref{eq:mle:usge_cf}-\ref{eq:mle:usge_b}) and omitting constants, we obtain an approximate solution for the optimization problem (\ref{eq:mle:usge_cf}), referred to as the Unknown-Signal Approximated Gaussian Estimator (USAGE) and given by
\begin{equation*}
    {\hat{\boldsymbol{q}}}_{USAGE}=\arg{\max\limits_{\boldsymbol{q}}\ {C_1(\boldsymbol{q})}},
\end{equation*} 
\begin{equation} \label{eq:usage:step2:usage_cf}
    C_1(\boldsymbol{q})\triangleq\max\limits_{\boldsymbol{\gamma}\in\mathbb{T}^{KD}}\ \left({\boldsymbol{\gamma}}^\ast\right)^\dag \boldsymbol{A}(\boldsymbol{q}){\boldsymbol{\gamma}}^\ast, 
\end{equation}
where 
\begin{equation}
    \begin{split}
        \boldsymbol{A}(\boldsymbol{q})\triangleq\sigma_v^{-4}&{\boldsymbol{\hat{\Gamma}}}\Biggl\{\sum_{m=0}^{M-1}{\boldsymbol{Y}_m^\dag\boldsymbol{G}_m(\boldsymbol{q})\boldsymbol{U}_m\Bigl[\boldsymbol{I}+\ldots} \\
        &+\sigma_v^{-2}\boldsymbol{U}_m^\dag\hat{\boldsymbol{\Gamma}}^2\boldsymbol{U}_m\Bigr]^{-1}\boldsymbol{U}_m^\dag\boldsymbol{G}_m^\dag(\boldsymbol{q})\boldsymbol{Y}_m\Biggr\}\boldsymbol{\hat{\Gamma}},
    \end{split} \label{eq:usage:step2:usage_A}
\end{equation}
\begin{gather*}
    \boldsymbol{Y}_m\triangleq\left[\boldsymbol{Y}^{0}_m,\ldots,\boldsymbol{Y}^{D-1}_m\right], \\
    \boldsymbol{Y}^{d}_m\triangleq diag\left\{\boldsymbol{y}^{d}_m\right\},
\end{gather*}
and
\begin{equation*}
    \mathbb{T}^N\triangleq\left\{\boldsymbol{\gamma}\in\mathbb{C}^N:\ \left|\gamma[0]\right|=...=\left|\gamma[N-1]\right|=1\right\}.
\end{equation*}

In (\ref{eq:usage:step2:usage_cf}) we have a Quadratic form maximization over the vector $\boldsymbol{\gamma}$, with elements that are constrained to unit magnitude. Notice that any solution, $\hat{\boldsymbol{\gamma}}$, is unique up to a constant phase shift. Namely, (\ref{eq:usage:step2:usage_cf}) has $KD-1$ degrees of freedom. To solve this non-convex optimization problem we use the low complexity Generalized Power Method (GPM) \cite{liu2017gpm} presented in Appendix \ref{append:gpm}. Algorithm \ref{algo:usage} summarizes the step-by-step operation of the proposed USAGE.

We note that other optimization methods for (\ref{eq:usage:step2:usage_cf}) were considered, such as PhaseCut \cite{waldspurger2015phase}, gradient-descent (applied directly over the phase vector, $\angle{\boldsymbol{x}}$), and more. The other methods had considerably higher computational complexity than GPM, while none achieved substantial performance advantage in position estimation.

In the previously mentioned case where the magnitudes are known, $\boldsymbol{\Gamma}$ does not need to be estimated in the first stage of the proposed algorithm, i.e., step 4 of Algorithm \ref{algo:usage}. Note that in that case, the optimization problem of USAGE (\ref{eq:usage:step2:usage_cf}-\ref{eq:usage:step2:usage_A}) coincides with that of the original optimization problem (\ref{eq:mle:usge_cf}-\ref{eq:mle:usge_b}).

\begin{algorithm}[H]
\caption{Unknown-Signal Approximated Gaussian Estimator (USAGE)}
\begin{algorithmic}[1]\label{algo:usage}
 \renewcommand{\algorithmicrequire}{\textbf{Parameters:}}
 \REQUIRE BSs positions $\left\{\boldsymbol{p_m}\right\}_{m=0}^{M-1}$, channel covariance matrices $\left\{\boldsymbol{H_m}\right\}_{m=0}^{M-1}$, number of windows $D$, GPM step size $\beta>0$
 \renewcommand{\algorithmicrequire}{\textbf{Input:}}
 \REQUIRE Frequency samples of the received signals $\left\{\boldsymbol{y}_m\right\}_{m=0}^{M-1}$, a set of potential transmitter positions $\left\{\boldsymbol{q}^n\right\}_{n=0}^{N_q-1}$
 \renewcommand{\algorithmicensure}{\textbf{Output:}}
 \ENSURE  Estimated transmitter position $\boldsymbol{\hat{q}}$
  \FOR {$m = 0$ to $M-1$}
  \STATE $\boldsymbol{U}_m \longleftarrow$ Eigen decomposition of $\boldsymbol{H_m}$ (see (\ref{eq:mle:ch_cov}))
  \ENDFOR
  \STATE Estimate magnitudes matrix $\boldsymbol{\hat{\Gamma}}$ using (\ref{eq:usage:step1:magest})
  \FOR {$n = 0$ to $N_q-1$}
  \FOR {$m = 0$ to $M-1$}
  \STATE Evaluate $\boldsymbol{G}_m$ for $\boldsymbol{q}^n$ and $\boldsymbol{p}_m$ using (\ref{eq:sysmodel:g_vec_def}-\ref{eq:mle:sig_freq_mat}) 
  \ENDFOR
  \STATE Evaluate $\boldsymbol{A}$ for $\boldsymbol{q}^n$ using (\ref{eq:usage:step2:usage_A})
  \STATE $\boldsymbol{\hat{\gamma}}^\ast\longleftarrow$ Run GPM (Algorithm \ref{algo:gpm}) with $\beta$ over $\boldsymbol{A}$
  \STATE $C_1[n] \longleftarrow \left({\boldsymbol{\hat{\gamma}}}^\ast\right)^\dag \boldsymbol{A}{\boldsymbol{\hat{\gamma}}}^\ast$
  \ENDFOR
  \STATE $\boldsymbol{\hat{q}} \longleftarrow \arg \max\limits_{\boldsymbol{q}^n}\ C_1[n]$
 \RETURN $\boldsymbol{\hat{q}}$
 \end{algorithmic} 
\end{algorithm}

\subsection{Optimization Complexity Reduction} \label{sec:usage:optcomp}
The computational complexity of GPM (Algorithm \ref{algo:gpm}), employed for solving (\ref{eq:usage:step2:usage_cf}), grows with the length of the received signal, $\boldsymbol{y}_m$, which is $KD$, where $D$ represents the number of observation windows and $K$ indicates the number of frequency samples (i.e., DFT coefficients) within each window. In this section, we aim to mitigate the complexity of the optimization problem in (\ref{eq:usage:step2:usage_cf}) by reducing the length of the signals given as input to USAGE by a factor of $D$.

Let us consider a phase-shift operation applied uniformly to the received signals from all BSs. Given any vector of phases, $\boldsymbol{\theta}\in\mathbb{R}^{KD}$, the operation is defined as follows
\begin{equation} \label{eq:usage:optcomp:y_tilde_def}
    \forall m:\ \Tilde{\boldsymbol{y}}_m\left(\boldsymbol{\theta}\right)\triangleq exp\left\{-j\boldsymbol{\theta}\right\}\odot\boldsymbol{y}_m,
\end{equation}
where $\odot$ is the Hadamard product. Substituting (\ref{eq:usage:optcomp:y_tilde_def}) into (\ref{eq:mle:sig_freq_mat}), we get
\begin{equation} \label{eq:usage:optcomp:y_tilde_model}
    \Tilde{\boldsymbol{y}}_m=\Tilde{\boldsymbol{X}}\boldsymbol{G}_m\boldsymbol{\eta}_m+\Tilde{\boldsymbol{v}}_m,
\end{equation}
where
\begin{gather*}
    \Tilde{\boldsymbol{v}}_m\left(\boldsymbol{\theta}\right)\triangleq exp\left\{-j\boldsymbol{\theta}\right\}\odot\boldsymbol{v}_m,\\
    \Tilde{\boldsymbol{x}}\left(\boldsymbol{\theta}\right)\triangleq exp\left\{-j\boldsymbol{\theta}\right\}\odot\boldsymbol{x},
\end{gather*}
and $\Tilde{\boldsymbol{X}}$ is the matrix representation of $\Tilde{\boldsymbol{x}}$, corresponding to the relation between $\boldsymbol{X}$ and $\boldsymbol{x}$.

Note that $\Tilde{\boldsymbol{v}}_m$ maintains the properties of a complex AWGN and is independent of $\boldsymbol{\theta}$. Furthermore, $\Tilde{\boldsymbol{x}}$ is still an unknown complex vector to be estimated. Thus, we conclude that our proposed estimator is invariant to a uniform phase-shift (much like other TDOA-based estimators), i.e., plugging $\Tilde{\boldsymbol{y}}_m$ into (\ref{eq:mle:usge_cf}-\ref{eq:mle:usge_b}) will yield the same cost-function value for any $\boldsymbol{\theta}$.

Next, let $\Delta\boldsymbol{\phi}^d\in\mathbb{R}^K$ denote the difference in phases of the transmitted signal between the $d$th and first observation windows, namely
\begin{equation} \label{eq:usage:optcomp:delta_phi_d}
    \Delta\boldsymbol{\phi}^d\triangleq\angle{\boldsymbol{x}^d}-\angle{\boldsymbol{x}^0},
\end{equation}
and let $\Delta\boldsymbol{\phi}\triangleq\begin{bmatrix}{\Delta\boldsymbol{\phi}^0}^T,&\ldots,&{\Delta\boldsymbol{\phi}^{D-1}}^T\end{bmatrix}^T$.

Assuming for now that $\Delta\boldsymbol{\phi}$ is known, we use it to apply a phase-shift. Consequently, the phases of the equivalent transmitted signal, $\Tilde{\boldsymbol{x}}\vert _{\theta=\Delta\phi}$, in different observation windows have all been aligned to those of the first window, i.e., $\angle{\Tilde{\boldsymbol{x}}^d}=\angle{\boldsymbol{x}^0}$ for any $d$. Thus, when using USAGE over $\left\{\Tilde{\boldsymbol{y}}_m\vert _{\theta=\Delta\phi}\right\}_{m=0}^{M-1}$, the optimization problem given in (\ref{eq:usage:step2:usage_cf}) is reduced to just $K-1$ degrees of freedom instead of $KD-1$. Furthermore, we can preserve the Quadratic form of the problem and use GPM to solve it by substituting the following
\begin{equation}
    \Tilde{\boldsymbol{\gamma}}\vert _{\theta=\Delta\phi}=\left[\underbrace{{\boldsymbol{\gamma}^0}^T,\ldots,{\boldsymbol{\gamma}^0}^T}_\text{D times}\right]^T=[\underbrace{\boldsymbol{I},\ldots,\boldsymbol{I}}_\text{D times}]^T\cdot\boldsymbol{\gamma}^0,
\end{equation}
into (\ref{eq:usage:step2:usage_cf}) and maximize over $\boldsymbol{\gamma}^0$, where $\boldsymbol{\gamma}^0\triangleq exp\left\{j\angle{\boldsymbol{x}^0}\right\}$.

To reduce the computational complexity even further, we replace $\Tilde{\boldsymbol{y}}_m\vert _{\theta=\Delta\phi}\in\mathbb{C}^{KD}$ with the sum of its windows, thus lowering the complexity of many matrix operations. Note that the summation is done coherently since the phases of all windows are now aligned (up to some measurement noise). Let $\Bar{\boldsymbol{y}}_m\in\mathbb{C}^K$ denote the coherent sum of the aligned observation windows at the $m$th BS
\begin{equation} \label{eq:usage:optcomp:y_hat_def}
    \Bar{\boldsymbol{y}}_m\triangleq\sum_{d=0}^{D-1}{\Tilde{\boldsymbol{y}}_m^d\vert _{\theta=\Delta\phi}}.
\end{equation}
Substituting (\ref{eq:usage:optcomp:delta_phi_d}-\ref{eq:usage:optcomp:y_hat_def}) into (\ref{eq:mle:sig_freq_mat}), we get
\begin{equation} \label{eq:usage:optcomp:y_hat_model}
    \Bar{\boldsymbol{y}}_m=\Bar{\boldsymbol{X}}\boldsymbol{G}_m\boldsymbol{\eta}_m+\Bar{\boldsymbol{v}}_m,
\end{equation}
where
\begin{gather*}
    \Bar{\boldsymbol{v}}_m\triangleq\sum_{d=0}^{D-1}{exp\left\{-j\Delta\boldsymbol{\phi}^d\right\}\odot\boldsymbol{v}_m^d},\\
    \Bar{\boldsymbol{x}}\triangleq \boldsymbol{\gamma}^0\odot\sum_{d=0}^{D-1}{\left|\boldsymbol{x}^d\right|},
\end{gather*}
and $\Bar{\boldsymbol{X}}$ is the matrix representation of $\Bar{\boldsymbol{x}}$, corresponding to the relation between $\boldsymbol{X}$ and $\boldsymbol{x}$.

Up to this point, we assumed $\Delta\boldsymbol{\phi}$ to be known, though this is generally not true. Hence, we now offer an iterative estimation method for both $\Delta\boldsymbol{\phi}$ and $\Bar{\boldsymbol{y}}_m$. Let $\Delta\Hat{\boldsymbol{\phi}}^d$ denote the estimated phase differnce vector at the $d$th window, and $\Hat{\bar{y}}_m^d$ represent the sum of $\Tilde{\boldsymbol{y}}_m^d\vert _{\theta=\Delta\Hat{\phi}^d}$ up to the $d$th window (inclusive), which is given by
\begin{equation} \label{eq:usage:optcomp:ybar_iter}
    \Hat{\Bar{\boldsymbol{y}}}_m^d\triangleq
    \begin{cases}
        exp\left\{-j\Delta\Hat{\boldsymbol{\phi}}^d\right\}\odot \boldsymbol{y}_m^{d} + \Hat{\Bar{\boldsymbol{y}}}_m^{d-1}, & d>0\\
        \boldsymbol{y}_m^{0}, & d=0
    \end{cases}.
\end{equation}
Notably, $\Delta{\boldsymbol{\phi}}^d$ is independent of $m$ and in a noiseless environment could be extracted from a single BS, as given by (\ref{eq:sysmodel:sig_freq}) and (\ref{eq:usage:optcomp:delta_phi_d}) 
\begin{equation*}
    \begin{split}
        \left(\boldsymbol{v}_m=\boldsymbol{0}\right)\Longrightarrow\angle\left\{\boldsymbol{y}^{d}_m\odot \left(\boldsymbol{y}^{0}_m\right)^\ast\right\}=\angle{\boldsymbol{x}^d}-\angle{\boldsymbol{x}^0}=\Delta{\boldsymbol{\phi}}^d.
    \end{split}
\end{equation*}
Nevertheless, to reduce the effects of the AWGN, we offer to incorporate measurements from all BSs in the following manner
\begin{equation} \label{eq:usage:optcomp:phase_iter}
    \Delta\Hat{\boldsymbol{\phi}}^{d}\triangleq
    \begin{cases}
        \angle\left\{\frac{1}{M}\sum_{m=0}^{M-1}{\boldsymbol{y}^{d}_m\odot \left(\Hat{\Bar{\boldsymbol{y}}}^{d-1}_m\right)^\ast}\right\}, & d>0\\
        \boldsymbol{0}, & d=0
    \end{cases}.
\end{equation}
A more in-depth discussion about the convergence of (\ref{eq:usage:optcomp:phase_iter}) into (\ref{eq:usage:optcomp:delta_phi_d}) and the needed SNR and BSs conditions is given in Appendix \ref{append:cwcderiv}.

Algorithm \ref{algo:cwc} outlines the iterative Coherent Window Combining (CWC) method, based on (\ref{eq:usage:optcomp:ybar_iter}-\ref{eq:usage:optcomp:phase_iter}). Its output is the reduced measurement vector, as defined in (\ref{eq:usage:optcomp:y_hat_def}).

\begin{algorithm}[H]
\caption{Coherent Window Combining (CWC)}
\begin{algorithmic}[1]\label{algo:cwc}
 \renewcommand{\algorithmicrequire}{\textbf{Input:}}
 \renewcommand{\algorithmicensure}{\textbf{Output:}}
 \REQUIRE $\left\{\boldsymbol{y}_m^d\right\}_{(m,d)=(0,0)}^{(M-1,D-1)}$ - DFT of received signals from multiple BSs and observation windows, where $\boldsymbol{y}_m^d\in\mathbb{C}^{K}$
 \ENSURE $\left\{\boldsymbol{\bar{y}}_m\right\}_{m=0}^{M-1}$ - Coherently combined DFTs of multiple BSs, where $\boldsymbol{\bar{y}}_m\in\mathbb{C}^{K}$
  \STATE $\forall m:\ \bar{\boldsymbol{y}}_m^0 \longleftarrow \boldsymbol{y}_m^0$
  \FOR {$d = 1$ to $D-1$}
  \STATE $\Delta\boldsymbol{\phi}^{d} \longleftarrow \angle\left\{\frac{1}{M}\sum_{m=0}^{M-1}{\boldsymbol{y}^{d}_m\odot \left(\bar{\boldsymbol{y}}^{d-1}_m\right)^\ast}\right\}$
  \STATE $\forall m:\ \bar{\boldsymbol{y}}_m^d \longleftarrow \bar{\boldsymbol{y}}_m^{d-1} + exp\left\{-j\Delta\boldsymbol{\phi}^{d}\right\}\odot \boldsymbol{y}_m^{d}$
  \ENDFOR
 \RETURN $\left\{\boldsymbol{\bar{y}}_m^{D-1}\right\}_{m=0}^{M-1}$
 \end{algorithmic} 
\end{algorithm}

In conclusion, given the appropriate amount of BSs and SNR conditions, we can apply CWC over the received signal measurements while preserving the system model as presented in Section \ref{sec:sysmod}. Thus, instead of employing USAGE (Algorithm \ref{algo:usage}) directly on the received measurements, we can employ it on the output of CWC, reducing its computational cost considerably, as demonstrated in Section \ref{sec:results:general}. We refer to this more efficient estimator as USAGE-CWC.
%

\section{Performance Analysis} \label{sec:crlb}
We turn to derive the CRLB for the position estimation problem presented in this paper under the Gaussian approximation of the channel, as discussed in Section \ref{sec:mle}. Thus, for a given transmitter position, $\boldsymbol{q}$, and transmitted signal, $\boldsymbol{x}$, the measurements vectors, $\left\{\boldsymbol{y}_m\right\}_{m=0}^{M-1}$, are independent complex Gaussian random vectors with zero mean and a set of covariance matrices $\left\{\boldsymbol{R}_m\right\}_{m=0}^{M-1}$. The element in the $u$th row and $v$th column of the Fisher information matrix (FIM) of such a set of independent complex Gaussian vectors with unknown real parameters vector $\boldsymbol{\xi}$ is given by \cite{BOOK:schreier2010performance}  
\begin{equation}\label{eq:crlb:fim_def}
    J_F\left[u,v\right] \triangleq \sum_{m=0}^{M-1}{Tr\left\{\boldsymbol{R}_m^{-1}\left(\frac{\partial\boldsymbol{R}_m}{\partial\xi_u}\right)\boldsymbol{R}_m^{-1}\left(\frac{\partial\boldsymbol{R}_m}{\partial\xi_v}\right)\right\}},
\end{equation}
where $\boldsymbol{J_F}$ is the FIM, $Tr\{\cdot\}$ is the matrix trace operator and the notation $\partial/\partial\xi_u$ indicates a partial derivative by $\xi_u$, which is the $u$th component of vector $\boldsymbol{\xi}$.

In the presented estimation problem the unknown parameters are the three-dimensional position coordinates of the transmitter, $\boldsymbol{q}=[q_x,q_y,q_z]^T$, and the complex transmitted signal, $\boldsymbol{x}\in\mathbb{C}^{KD}$, which is decomposed to magnitudes vector, $\left|\boldsymbol{x}\right|$, and phases vector, $\angle\boldsymbol{x}$. As mentioned in Section \ref{sec:usage:step2}, the solution for $\boldsymbol{x}$ is unique up to a constant phase shift, thus, only $KD-1$ phase elements need to be estimated. And so, the vector of unknown real parameters is denoted by
 \begin{equation} \label{eq:crlb:paramset}
    \boldsymbol{\xi}=\left[\boldsymbol{q}^T,\left|\boldsymbol{x}\right|^T,\angle{x[0]},\ldots,\angle{x[KD-2]}\right]^T.
 \end{equation}

In order to evaluate the FIM and CRLB, we first derive explicit expressions for ${\partial\boldsymbol{R}_m}/{\partial\xi_u}$. Starting with the covariance matrix derivative with respect to the unknown position parameters, we plug (\ref{eq:mle:toa}) into (\ref{eq:mle:cov}) and differentiate with respect to each coordinate to get
\begin{gather}
\begin{split}
    \frac{\partial\boldsymbol{R}_m}{\partial q_{x,y,z}}=\boldsymbol{X}\left(\frac{\partial\boldsymbol{G}_m}{\partial q_{x,y,z}}\right)\boldsymbol{H}_m\boldsymbol{G}_m^\dag\boldsymbol{X}^\dag\ldots\\
    +\boldsymbol{X}\boldsymbol{G}_m\boldsymbol{H}_m\left(\frac{\partial\boldsymbol{G}_m}{\partial q_{x,y,z}}\right)^\dag\boldsymbol{X}^\dag,
\end{split} \label{eq:crlb:cov_div_q} \\
\frac{\partial\boldsymbol{G}_m}{\partial q_{x,y,z}}=-j\frac{2\pi}{c}\frac{ \left(\boldsymbol{q}-\boldsymbol{p}_m\right)_{x,y,z}}{\left\|\boldsymbol{q}-\boldsymbol{p}_m\right\|}\cdot diag\left\{\boldsymbol{f}\right\} \cdot \boldsymbol{G}_m, \label{eq:crlb:G_div_q}
\end{gather}
where $\boldsymbol{f}\triangleq\left[f_0,\ldots,f_{K-1}\right]^T$.

Moving on to the derivative with respect to the unknown signal magnitude parameters, recall that $x[k+Kd]=x^d[k]$. From (\ref{eq:mle:cov}) together with the definition of $\boldsymbol{X}$ we get
\begin{equation}\label{eq:crlb:cov_div_x_abs}
\begin{split}
    \frac{\partial\boldsymbol{R}_m}{\partial |x^d[k]|}=e^{j\angle{x^d[k]}}\boldsymbol{\Phi_k^d}\boldsymbol{G}_m\boldsymbol{H}_m\boldsymbol{G}_m^\dag\boldsymbol{X}^\dag\ldots\\
    +e^{-j\angle{x^d[k]}}\boldsymbol{X}\boldsymbol{G}_m\boldsymbol{H}_m\boldsymbol{G}_m^\dag\left(\boldsymbol{\Phi_k^d}\right)^T,
\end{split}
\end{equation}
where $\boldsymbol{\Phi_k^d}\in\mathbb{R}^{KD\times K}$ and its element at the $u$th row and $v$th column is given by
\begin{gather*}
    \Phi_k^d[u,v]\triangleq\delta[k+Kd-u]\cdot \delta[k-v],
\end{gather*}
with $\delta[\cdot]$ representing the Kronecker delta function. Lastly, the covariance matrix derivative with respect to the unknown signal phase parameters is given by
\begin{equation}\label{eq:crlb:cov_div_x_phase}
\begin{split}
    \frac{\partial\boldsymbol{R}_m}{\partial \angle x^d[k]}=jx^d[k]\boldsymbol{\Phi_k^d}\boldsymbol{G}_m\boldsymbol{H}_m\boldsymbol{G}_m^\dag\boldsymbol{X}^\dag\ldots\\
    -j{x^d[k]}^\ast\boldsymbol{X}\boldsymbol{G}_m\boldsymbol{H}_m\boldsymbol{G}_m^\dag\left(\boldsymbol{\Phi_k^d}\right)^T.
\end{split}
\end{equation}
By substituting (\ref{eq:mle:cov_inv}) and (\ref{eq:crlb:paramset}-\ref{eq:crlb:cov_div_x_phase}) into (\ref{eq:crlb:fim_def}) we get the FIM for our estimation problem.

Let $\boldsymbol{\Sigma}_{\boldsymbol{\xi}}$ denote the CRLB matrix for the estimation of $\boldsymbol{\xi}$, then given (\ref{eq:crlb:paramset}) it satisfies
\begin{equation} \label{eq:crlb:crlb_mat}
    \boldsymbol{\Sigma}_{\boldsymbol{\xi}}=\boldsymbol{J_F}^{-1}=\begin{bmatrix}\boldsymbol{\Sigma}_{\boldsymbol{q}}&\dotsc\\\dotsc&\dotsc\end{bmatrix},
\end{equation}
where $\boldsymbol{\Sigma}_{\boldsymbol{q}}\in\mathbb{R}^{3\times 3}$ is a sub-matrix of $\boldsymbol{\Sigma}_{\boldsymbol{\xi}}$ corresponding to the covariance matrix of the three-dimensional position estimation error. The CRLB for the emitter's position estimation is then given by
\begin{equation}\label{eq:crlb:mse_q}
    \sigma_q^2=Tr\left\{\boldsymbol{\boldsymbol{\Sigma}}_{\boldsymbol{q}}\right\},
\end{equation}
\begin{equation*}
    E\left\{\left\|\boldsymbol{q}-\hat{\boldsymbol{q}}\right\|^2\right\}\geq \sigma_q^2,
\end{equation*}
where $\hat{\boldsymbol{q}}$ is the estimated position.

As mentioned, there are scenarios where the signal magnitudes are known a priori. In that case, we set $\boldsymbol{\xi}=\left[\boldsymbol{q}^T,\angle{x[0]},\ldots,\angle{x[KD-2]}\right]^T$ and obtain the appropriate FIM and CRLB in the same manner, leaving out the magnitude-related elements.

\section{Results and Discussion} \label{sec:results}
In this section, we present simulated performance results of USAGE and USAGE-CWC in various multipath channel models, with different types of transmit signals, a range of SNR levels, and a varying number of BSs. We compared the performance of our proposed estimators to each other and to three reference estimators, taken from the literature.

The section is organized as follows: Section \ref{sec:results:refposest} contains a brief explanation of each of the three reference estimators. In Section \ref{sec:results:simprop} we give an overview of simulation properties and configurations (e.g., channel model and transmit signal properties, scattering of BSs, etc.). The performance results of the estimators when the transmitted signal phases and magnitudes are unknown are presented in Section \ref{sec:results:general}, and in Section \ref{sec:results:special} we compare the performance of USAGE-CWC when the signal magnitudes are known vs. unknown (as in the preceding section).

\subsection{Reference Estimators} \label{sec:results:refposest}
The first reference estimator we consider is the position MLE developed for scenarios involving an unknown signal in a single LOS path channel \cite{MUSIC:tirer2016high}. We will refer to this estimator as SML.

The second reference estimator we examine is based on Signal-Subspace Projection MUSIC (SSP-MUSIC) \cite{MUSIC:du2018music}. SSP-MUSIC is a position estimator designed for scenarios involving an unknown transmitted signal in a man-made multipath environment where transponders at known locations are utilized. In this setting, the arrival times of NLOS paths relative to the LOS path depend on the transmitter's position. However, in the system model presented in this paper, the locations of reflecting objects are unknown, resulting in random relative arrival times. To adapt SSP-MUSIC to our model, we constrained the arrival times of the paths considered by the estimator to a finite grid, i.e., $\tau_{m,l}'=l\cdot\Delta\tau$, where $\Delta\tau$ is an adjustable parameter. The modified SSP-MUSIC (MSSP-MUSIC), estimates the complex gain corresponding to each arrival time on the defined grid. The grid spacing and the number of paths considered were optimized by simulations for each channel model employed in our study.

The third and final reference estimator, introduced in a recent work by Kehui \textit{et al.} \cite{zhu2022cross}, is referred to as the Cross-Spectra MUSIC (CS-MUSIC) estimator. CS-MUSIC is designed to estimate the position of a transmitter in a system where both the transmitted signal and reflecting objects are unknown, similar to our system model. The CS-MUSIC estimator operates by applying the MUSIC algorithm to the DFT of the cross-correlation between the received signal at a reference BS and other BSs. The specific parameters governing the CS-MUSIC estimator were optimized for each channel model.

Concerning the parameters of our proposed estimators, namely USAGE and USAGE-CWC, we determined the channel covariance matrix, $\boldsymbol{H}_m$, using the PDP of each channel model, which was measured empirically from $10^3$ simulated channel instances. The noise variance, $\sigma_v^2$, is assumed to be known as it can be obtained independently from measurements before or between transmissions.

Note that all of the aforementioned estimators employ partitioning of the full observation time into smaller intervals, as utilized in the derivation of our proposed estimator. To ensure a fair comparison among these estimators, we will set identical observation window configurations.

\subsection{Simulation Setup} \label{sec:results:simprop}
The performance of the proposed and reference estimators was evaluated by Monte Carlo simulations. For each evaluation, 10 different position configurations of the transmitter and BSs were taken at random (uniformly). For each configuration, we ran 100 estimation trials. At each trial, different channels and transmitted signal realizations were randomly generated. Our performance metric is the root-mean-square error (RMSE) for position estimation across 1000 trials, defined as follows
\begin{equation} \label{eq:results:simsetup:rmse}
    RMSE = \sqrt{E\left\{\left\|\boldsymbol{q}-\hat{\boldsymbol{q}}\right\|^2\right\}}
\end{equation}

The scalar stochastic CRLB \cite{BOOK:schreier2010performance} that corresponds to the above RMSE was obtained by using (\ref{eq:crlb:mse_q}) with the average FIM (averaged over the 1000 simulated trials).

In our simulations, we used two Rayleigh fading channel models with an exponential decaying PDP denoted by 'Exp1' and 'Exp2'. In these models, the multipath coefficients, $\alpha_{m,l}$, are independent and have zero-mean complex Gaussian distribution, and the multipath delays are known. Thus, these channels perfectly match the Gaussian channel assumption under which our proposed estimator was derived. Each channel is made out of $L$ paths with a constant time spacing, denoted by $\Delta\tau$, where $\Delta\tau \ll W^{-1}$, and the paths' gains decay exponentially with the delay. The gain variance of the $l$th path is given by
\begin{equation}\label{eq:results:simsetup:ch_exp_var}
    \sigma_{m,l}^2={\mu_0}^{los}\delta[l]+{\mu_0}^{nlos} e^{-\frac{l\Delta\tau}{\mu_1}}u[l-1],
\end{equation}
where $\mu_0^{los}\ge 0$ and $\mu_0^{nlos}\ge 0$ are measures of power for the LOS and NLOS paths respectively, $\mu_1>0$ is the inverse of the NLOS gain exponential decay rate, and $u[l]$ is the discrete unit step function. 
Unless mentioned otherwise, the parameters used for Exp1 and Exp2 were as presented in Table \ref{tab:results:simsetup:ch_exp_param}.

An additional channel model employed in our simulations is the fifth UWB channel model, originally proposed by the IEEE 802.15.4a study group \cite{uwbchan3} and designed to replicate outdoor suburban environments. This particular model is referred to as 'UWB5'.

\begin{table}[!b]
    \renewcommand{\arraystretch}{1.3}
    \centering
    \begin{tabular}{c|c c c c c}
        \hline
        \renewcommand{\extrarowheight}{1cm}
        \textbf{Ch. Model}&$\boldsymbol{\mu^{los}_0}$&$\boldsymbol{\mu^{nlos}_0}$&$\boldsymbol{\mu_1}$ [ns]&$\boldsymbol{\Delta\tau}$ [ns]&$\boldsymbol{L}$\\
         \hline \hline
        Exp1 & 0.45 & 0.1 & 20 & 1 & 100\\
        Exp2 & 0.098 & 0.13 & 30 & 1 & 300\\
        \hline
    \end{tabular}
    \caption{Parameters of Exp1 and Exp2 channel models.}
    \label{tab:results:simsetup:ch_exp_param}
\end{table}

The transmitted signal was generated in the time domain. We tested two signal types, 'White' and 'Flat'. To generate a White signal we took $K$ random samples per window of a white complex Gaussian process. To generate a Flat signal we used a standard OFDM scheme with $K$ frequency bins (no pilots) per window. Each bin contained a 256-PSK modulated symbol. The Flat signal type can be an example of applying our proposed estimator on an OFDM system with constant amplitude modulation, since PSK symbols have a constant magnitude, i.e., $\forall (k,d): \left|x^d[k]\right|=const$. We assume that time synchronization with the OFDM symbols was achieved in this case. We used both signal types with BW of either 80 or 160 MHz.

\begin{figure}[!t]
    \centering
    \includegraphics[scale=0.4]{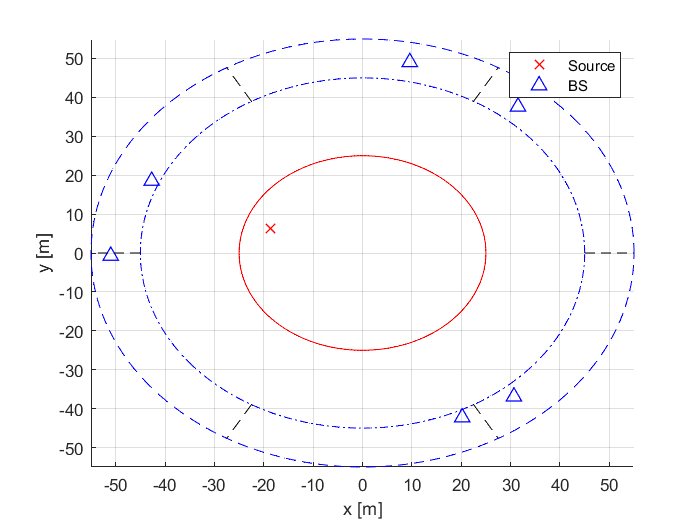}
    \caption{Example of positioning six BSs around a source in a simulation. The source was randomly positioned inside the red circle. The BSs were randomly positioned between the two blue-dashed lines in six different segments divided by the black-dashed lines.}
    \label{fig:simsetup_geo}
\end{figure}

The transmission source and BSs were positioned on a 2D plane as illustrated in Fig. \ref{fig:simsetup_geo}. In each configuration of source and BSs, the source was randomly positioned within a radius of 25 meters (red circle) and each BS was randomly placed in a separate segment. The segments had a radial range of 45-55 meters (blue dashed lines) and were divided into $M$ evenly spaced arcs (black dashed lines).

For the sake of computational load comparison, we state that simulations were written in MATLAB R2018a and executed on a computer with an Intel i7-1165G7 processor, 16 GB RAM, and Windows 11 Pro 64-bit operating system.

In the following sections, we test various system configurations. To simplify the reader's orientation we summarize in Table \ref{tab:results:sim_params} the main parameters used in the figures presented in the following. These parameters will also be further discussed in the following sections.

\begin{table*}[ht]
    \renewcommand{\arraystretch}{1.3}
    \centering
    \begin{tabular}{c c c c c c c c}
        \hline
        \renewcommand{\extrarowheight}{1cm}
        \textbf{Figure Index} & \textbf{Signal Type} & \textbf{Channel Model} & \textbf{BW} [MHz] & $\boldsymbol{K}$ & $\boldsymbol{D}$ & $\boldsymbol{M}$ & \textbf{SNR} [dB]\\
         \hline \hline
        \ref{fig:results:general:cwc} & White & Exp1 & 80 & 32 & 10 & 4$\sim$16 & 10$\sim$30\\
        \ref{fig:results:general:RMSEvsDS} & White & Exp2 (modified) & 160 & 64 & 5 & 12 & 25\\
        \ref{fig:results:general:RMSEvsM} & White & Exp2, UWB5 & 160 & 64 & 10 & 4$\sim$16 & 30\\
        \ref{fig:results:general:RMSEvsSNR} & White & Exp2, UWB5 & 160 & 64 & 10 & 16 & 10$\sim$30\\
        \ref{fig:results:special:RMSEvsM} & Flat & Exp2, UWB5 & 160 & 64 & 10 & 4$\sim$16 & 20\\
         \hline \\
    \end{tabular}
    \caption{Summarized simulation properties used in figures throughout Section \ref{sec:results}.}
    \label{tab:results:sim_params}
\end{table*}

\subsection{Position Estimation for Unknown Magnitudes} \label{sec:results:general}
In this section, we seek to assess the performance of our proposed estimator as well as the three reference estimators in terms of position estimation accuracy. The evaluation is conducted under the condition of a White signal transmission, wherein both the phases and magnitudes of the transmitted signal remain unknown.

\begin{figure}[!t]
\centering
\includegraphics[scale=0.4]{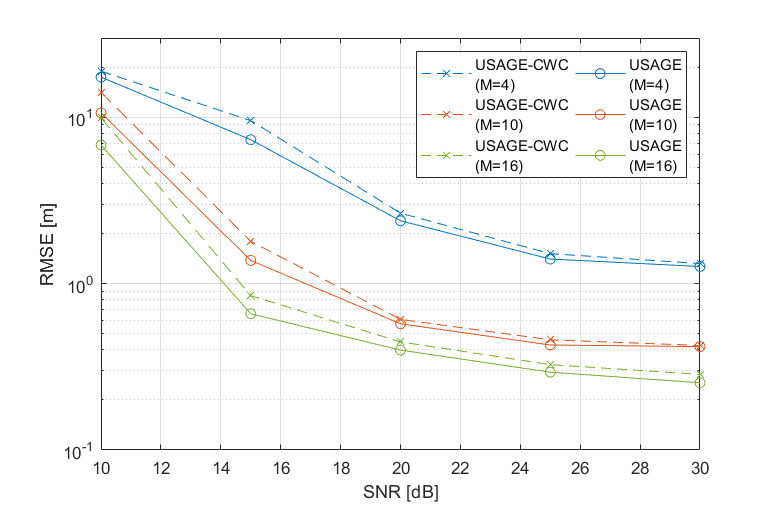}
\caption{RMSE vs. SNR comparison between USAGE and USAGE-CWC using the Exp1 channel model across varying numbers of BSs ($M$).}
\label{fig:results:general:cwc}
\end{figure}

Fig. \ref{fig:results:general:cwc} presents a comparative analysis of the performance between USAGE (Algorithm \ref{algo:usage}) and USAGE-CWC, which involves reducing the input dimension from $KD$ to $K$ by employing CWC (Algorithm \ref{algo:cwc}) before applying USAGE. The channel model utilized in this figure is Exp1, with a received signal BW of 80 MHz, while setting $K=32$ and $D=10$.

The results indicate that the performance disparity between USAGE and USAGE-CWC is generally small and diminishes with the increase of SNR. Regarding the runtime advantages of using CWC, the average runtime of USAGE in these simulations varied between 19 to 38 seconds per estimation trial (depending on $M$) while USAGE-CWC's runtime varied between 1.3 to 2.2 seconds per trial. These results conclusively indicate that incorporating CWC significantly reduces the computational load while maintaining the accuracy of USAGE for almost any practical use. Hence, from this point on we shall examine only USAGE-CWC due to its superior computational efficiency and comparable performance.


\begin{figure}[!t]
\centering
\includegraphics[scale=0.45]{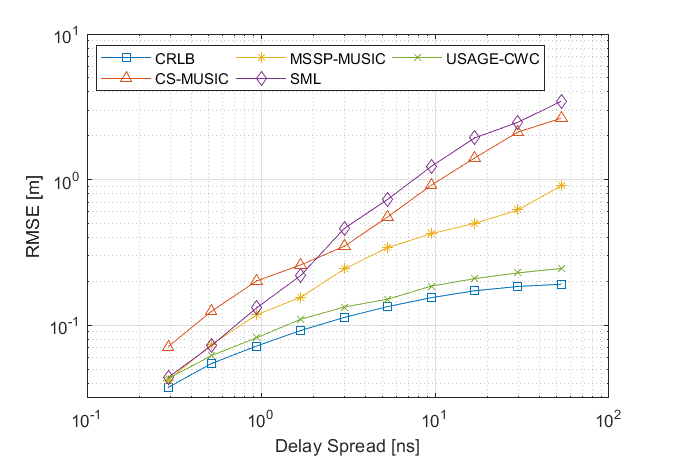}
\caption{RMSE vs. Delay-Spread of USAGE-CWC and reference estimators, in a modified Exp2 channel}
\label{fig:results:general:RMSEvsDS}
\end{figure}

Fig. \ref{fig:results:general:RMSEvsDS} illustrates the RMSE performance of USAGE-CWC, SML, MSSP-MUSIC, and CS-MUSIC as a function of the channel's delay spread, as well as the CRLB analysis. The transmitted signal in this evaluation is a White signal with a BW of 160 MHz. The graph employs a channel model based on Exp2, and to control the delay spread, we adjusted the parameter $\mu_1$ while ensuring that the ratio between the average energy of the NLOS and LOS components, given by $\frac{\mu_0^{nlos}}{\mu_0^{los}}\sum\limits_{l=1}^{L-1}{e^{-l\Delta\tau/\mu_1}}$, remains constant by adjusting $\mu_0^{nlos}$. The simulation was conducted with $M=12$ and an SNR of 25 dB. Notably, USAGE-CWC performs better than the other estimators for any delay spread, maintaining a relatively small gap from the CRLB.

Another noteworthy observation from Figure \ref{fig:results:general:RMSEvsDS} is the convergence of RMSE values of USAGE-CWC, SML, and MSSP-MUSIC to the CRLB as the channel's delay spread decreases. This convergence is expected since a channel with a delay spread shorter than the duration of a single transmit symbol can effectively be modeled as a single-path channel, which is exactly the model SML was developed for. Specifically, in Fig. \ref{fig:results:general:RMSEvsDS}, the delay spreads at which the estimators converge are shorter than 0.52 ns, considering a symbol period of around 6.3 ns, which is more than 10 times longer than the delay spread.

%
%

\begin{figure}[!t]
\centering
\subfloat[]{\includegraphics[scale=0.4]{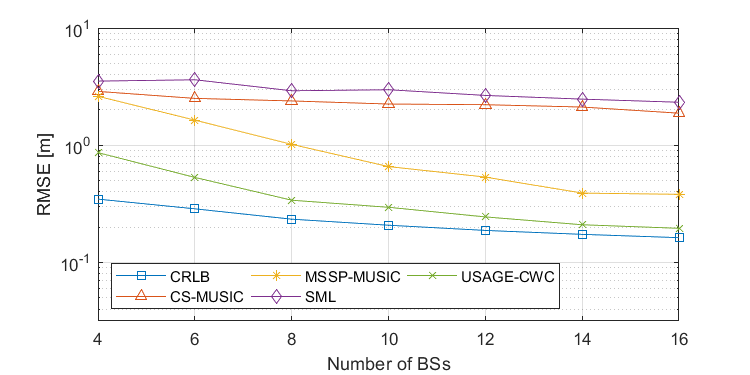}%
\label{fig:results:general:RMSEvsM:exp}}
\hfil
\subfloat[]{\includegraphics[scale=0.4]{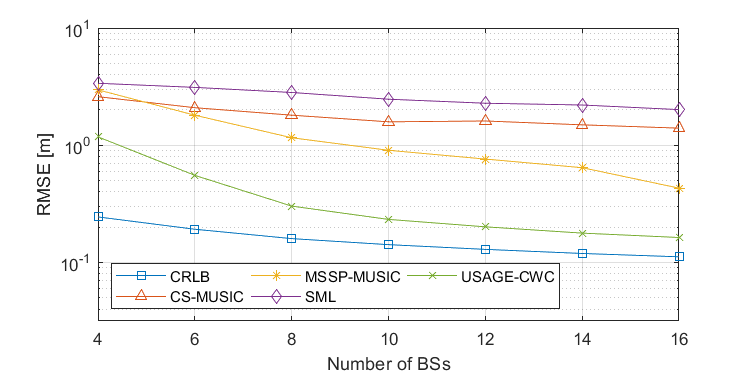}%
\label{fig:results:general:RMSEvsM:uwb5}}
\caption{RMSE vs. Number of BSs of USAGE-CWC and the three reference estimators for channel model: (a) Exp2 (b) UWB5.}
\label{fig:results:general:RMSEvsM}
\end{figure}

\begin{figure}[!t]
\centering
\subfloat[]{\includegraphics[scale=0.4]{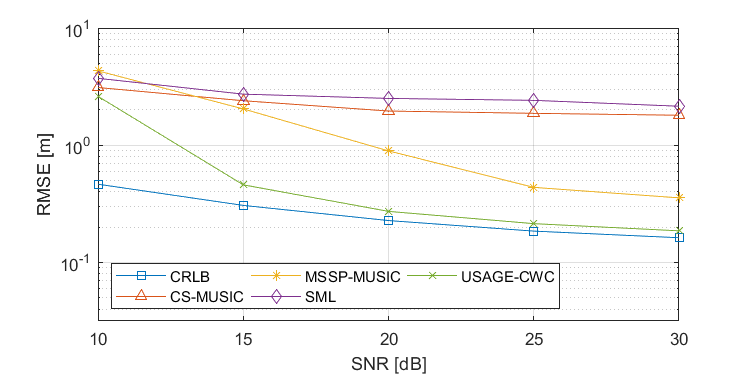}%
\label{fig:results:general:RMSEvsSNR:exp}}
\hfil
\subfloat[]{\includegraphics[scale=0.4]{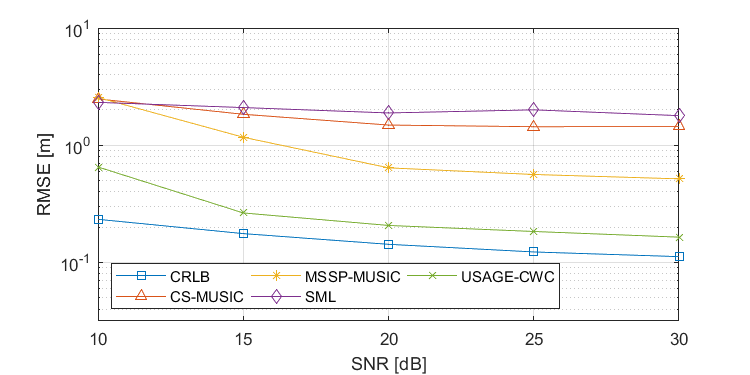}%
\label{fig:results:general:RMSEvsSNR:uwb5}}
\caption{RMSE vs. SNR of USAGE-CWC and the three reference estimators for channel model: (a) Exp2 (b) UWB5.}
\label{fig:results:general:RMSEvsSNR}
\end{figure}

We proceed to assess the performance of USAGE-CWC and the three reference estimators applied over both Exp2 and UWB5 channel models. The evaluation entails examining the variation in RMSE with respect to SNR and $M$. Fig. \ref{fig:results:general:RMSEvsM} illustrates the RMSE as a function of $M$ at a constant SNR of 30 dB, while Fig. \ref{fig:results:general:RMSEvsSNR} displays the RMSE against SNR for a fixed $M$ value of 16. In both figures, it is evident that USAGE-CWC surpasses the performance of the reference methods.

Fig. \ref{fig:results:general:RMSEvsM} and Fig. \ref{fig:results:general:RMSEvsSNR} reveal an important finding: in the Exp2 channel, USAGE-CWC approaches the CRLB as the number of BSs and SNR increase. However, when considering the UWB5 channel, a small but noticeable disparity persists, even at high values of BSs and SNR. To explain this phenomenon, we turn to the following explanation. In Section \ref{sec:mle}, the derivation of the likelihood function assumes a Gaussian nature for the received signal, and the same goes for the CRLB derivation in Section \ref{sec:crlb}. This assumption holds precisely when utilizing the Exp2 channel, as it generates a Gaussian channel. However, the UWB5 channel differs in that its received signal is only approximately Gaussian (due to the CLT) rather than being exactly Gaussian. Consequently, USAGE-CWC exhibits a slight performance gap from the CRLB in the UWB5 channel. Nevertheless, it should be noted that even in cases where the channel deviates from Gaussian, the Gaussian approximation provides an estimator with a commendable performance.

Recall that USAGE only solves an approximation of the original optimization problem (\ref{eq:mle:usge_cf}) since it estimates the signal magnitudes separately from the phases and position. The fact that USAGE-CWC achieves performance on par with the CRLB suggests that the approximations made during the derivation of USAGE hold true.

\subsection{Position Estimation for Known Magnitudes} \label{sec:results:special}
In this section, we focus on the performance of our proposed estimator in a specific scenario where the magnitudes of the transmitted signal are known. This evaluation is carried out under the condition of a Flat signal transmission, where the magnitudes of all frequency samples are equal and predetermined, while their phases remain unknown. Consequently, the CRLB presented throughout this section has been derived by treating the magnitudes as known quantities, as discussed in the final paragraph of Section \ref{sec:crlb}.

\begin{figure}[!t]
\centering
\subfloat[]{\includegraphics[scale=0.4]{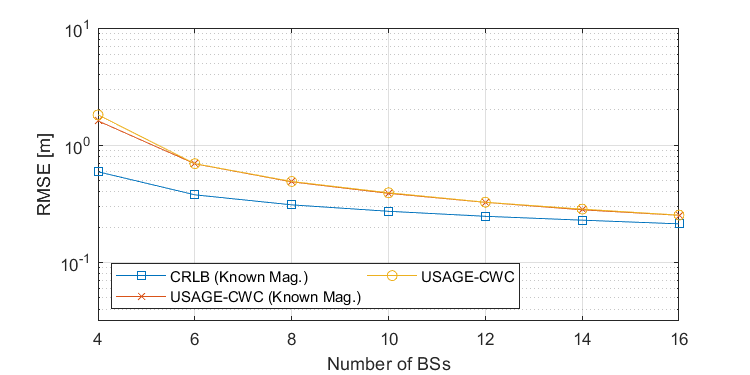}%
\label{fig:results:special:RMSEvsM:exp}}
\hfil
\subfloat[]{\includegraphics[scale=0.4]{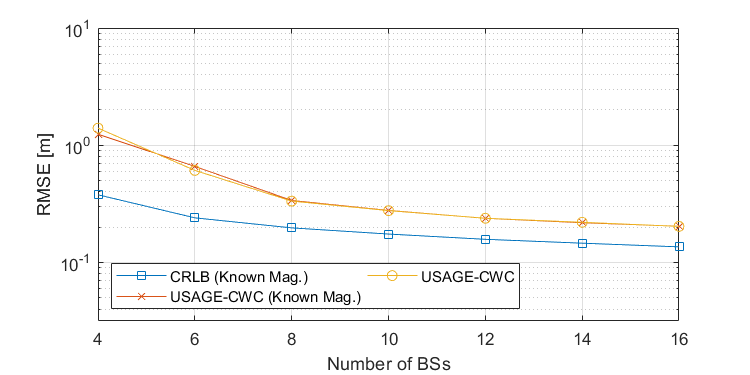}%
\label{fig:results:special:RMSEvsM:uwb5}}
\caption{RMSE vs. Number of BSs of USAGE-CWC under known and unknown magnitudes conditions, for channel model: (a) Exp2 (b) UWB5}
\label{fig:results:special:RMSEvsM}
\end{figure}

Fig. \ref{fig:results:special:RMSEvsM} offers a comparative analysis of the RMSE performance of USAGE-CWC in two scenarios: with and without knowledge of the transmit signal magnitudes. The number of BSs is varied, and both the Exp2 and UWB5 channel models are considered. The evaluation is conducted with an SNR of 20 dB. Interestingly, it is observed that leveraging the knowledge of the magnitudes has minimal impact on the RMSE. Additionally, a similar trend is observed in terms of approaching the CRLB, as illustrated in Fig. \ref{fig:results:general:RMSEvsM}, where a White signal at an SNR of 30 dB was employed.

As explained in Section \ref{sec:usage:step2}, when the magnitudes are known, the optimization problem for USAGE (expressed in (\ref{eq:usage:step2:usage_cf}-\ref{eq:usage:step2:usage_A})) aligns with that of the original problem (\ref{eq:mle:usge_cf}-\ref{eq:mle:usge_b}), under identical conditions. The performance of USAGE with unknown magnitudes is theoretically bounded by the MLE with unknown magnitudes, while the MLE with unknown magnitudes is bounded by the MLE with known magnitudes, which is equivalent to USAGE. Based on this rationale, observing that USAGE-CWC achieves the same RMSE results regardless of whether the magnitudes are known or unknown, we gather evidence that USAGE-CWC with unknown magnitudes approaches the MLE in cases involving unknown magnitudes. This serves as further validation for the approximations we employed during the derivation of USAGE.



\section{Conclusion} \label{sec:conclusion}
We have developed a novel position estimator for the localization of emitters transmitting signals of unknown characteristics within dense multipath environments. This estimator relies on prior knowledge of the PDP of the channel, which can be estimated beforehand. Additionally, we have derived a closed-form analytical expression for the CRLB in this study. In our investigations, the proposed estimator consistently outperformed other state-of-the-art estimators across two distinct simulated channel models, particularly at low to moderate SNR. Moreover, as SNR levels and the number of BSs increased, our estimator gradually approached the CRLB, affirming its accuracy and robustness. Furthermore, we introduced a complementary approach by integrating our estimator with the proposed CWC algorithm. This combination substantially reduced runtime with negligible performance degradation, showcasing its practical utility.


\appendices
\section{Generalized Power Method}\label{append:gpm}
The Generalized Power Method (GPM) is a modified iterative gradient method for solving the following optimization problem
\begin{equation}\label{eq:gpm:problem}
    \hat{\boldsymbol{\gamma}}=\arg \max\limits_{\boldsymbol{\gamma}\in\mathbb{T}^N}{\boldsymbol{\gamma}^\dag \boldsymbol{A}\boldsymbol{\gamma}}
\end{equation}
where $\boldsymbol{A}\in\mathbb{C}^{N\times N}$ is positive semi-definite (PSD).

The GPM algorithm used in this paper is presented in Algorithm \ref{algo:gpm}. It is based on the GPM algorithm presented in \cite{liu2017gpm}, with some notations adjustments.
\begin{algorithm}[H]
\caption{Generalized Power Method (GPM)}
\begin{algorithmic}[1]\label{algo:gpm}
 \renewcommand{\algorithmicrequire}{\textbf{Input:}}
 \renewcommand{\algorithmicensure}{\textbf{Output:}}
 \REQUIRE $\boldsymbol{A}$ - objective PSD matrix, $\beta>0$ - step size 
 \ENSURE $\boldsymbol{\hat{\gamma}}$ - an estimated solution for (\ref{eq:gpm:problem})
 \STATE $\boldsymbol{v}^{(0)} \longleftarrow$ leading eigenvector of $\boldsymbol{A}$
 \STATE $\boldsymbol{\gamma}^{(0)} \longleftarrow exp\left\{j\angle{\boldsymbol{v}^{(0)}}\right\}$
 \FOR {$i = 0,1,...$}
 \IF{termination criteria is met}
 \RETURN $\boldsymbol{\gamma}^{(i)}$
 \ELSE
 \STATE $\boldsymbol{v}^{(i+1)} \longleftarrow \left(\boldsymbol{I}+\beta\boldsymbol{A}\right)\boldsymbol{\gamma}^{(i)}$
 \STATE $\boldsymbol{\gamma}^{(i+1)} \longleftarrow exp\left\{j\angle{\boldsymbol{v}^{(i+1)}}\right\}$
 \ENDIF
 \ENDFOR
 \end{algorithmic} 
\end{algorithm}

For the purpose of this paper, we implemented termination criteria such that Algorithm \ref{algo:gpm} would stop when the cost-function, ${\boldsymbol{\gamma}^{(i)}}^\dag \boldsymbol{A}\boldsymbol{\gamma}^{(i)}$, of two consecutive iterations have a relative change smaller than $10^{-9}$, or after $10^{4}$ iterations. In our simulations, GPM gave similar empirical results for a wide range of step size values. We used $\beta\in\left[{10}^2,{10}^3\right]$ for the most part.

\section{Convergence of $\Delta\boldsymbol{\phi}$ Estimation in CWC}\label{append:cwcderiv}
In Section \ref{sec:usage:optcomp} we introduced CWC (Algorithm \ref{algo:cwc}) to be applied over samples given by (\ref{eq:mle:sig_freq_mat}). We note that CWC acts separately and unconditionally over each frequency bin, hence, for convenience purposes only we shall omit the frequency index, $k$, from the mathematical expressions to be followed. Thus, the scalar received sample is of the form
\begin{equation} \label{eq:cwcderiv:input}
    y_m^d = x^d g_{\tau_{m,0}} \eta_m + v_m^d
\end{equation}

Let us examine the first integration step ($d=1$) of CWC. Substituting (\ref{eq:cwcderiv:input}) and (\ref{eq:usage:optcomp:ybar_iter}) into (\ref{eq:usage:optcomp:phase_iter}), we get
\begin{equation} \label{eq:cwcderiv:phase_first}
\begin{split}
    \Delta\hat{\phi}^1 &= \angle\left\{\frac{1}{M}\sum_{m=0}^{M-1}{y^1_m\cdot {y^0_m}^\ast}\right\} \\
    &= \angle\left\{\psi^1_{xx}+\psi^1_{xv}+\psi^1_{vv}\right\},
\end{split}
\end{equation}
where
\begin{gather*}
    \psi^1_{xx}= x^1\cdot {x^0}^\ast\frac{1}{M}\sum_{m=0}^{M-1}{\left|\eta_m\right|^2},\\
    \psi^1_{xv}= \frac{1}{M}\sum_{m=0}^{M-1}{x^1g_{\tau_{m,0}}\eta_m{v_m^0}^\ast+\left(x^0 g_{\tau_{m,0}}\eta_m\right)^\ast{v_m^1}},\\
    \psi^1_{vv}= \frac{1}{M}\sum_{m=0}^{M-1}{v_m^1{v_m^0}^\ast}.
\end{gather*}

We turn to examine the behavior of (\ref{eq:cwcderiv:phase_first}) for a high number of BSs ($M\gg1$) and/or high SNR. Starting with the case of $M\gg1$, then according to the weak law of large numbers (WLLN) \cite{BOOK:mittelhammer2013mathematical}, the sample average of statistically independent random variables with finite mean converges in probability towards the expected value as the number of measurements grows. In our case, the number of measurements is the number of BSs. Thus, for $M\gg 1$ we approximate
\begin{gather*}
    \psi^1_{xx}\approx x^1{x^0}^\ast E\left\{\left|\eta_m\right|^2\right\},\\
    \begin{split}
        \psi^1_{xv}\approx\ &x^1E\left\{g_{\tau_{m,0}}\eta_m\right\} E\left\{{v_m^0}^\ast\right\}\ldots\\
        &+\left(x^0 \right)^\ast E\left\{g_{\tau_{m,0}}\eta_m^\ast\right\} E\left\{{v_m^1}\right\}=0,
    \end{split}\\
    \psi^1_{vv}\approx E\left\{v_m^1\right\} E\left\{{v_m^0}^\ast\right\}=0,
\end{gather*}
where the channel coefficients, $\eta_m$, and noise, $v_m^{0,1}$, are statistically independent with zero mean (as discussed in Section \ref{sec:mle}). We see that for a large enough $M$, the phase difference in the first CWC iteration can be approximated by
\begin{equation} \label{eq:cwcderiv:phase_first_approx}
    \Delta\hat{\phi}^1\approx \angle\left\{x^1{x^0}^\ast E\left\{\left|\eta_m\right|^2\right\}\right\}=\angle{x^1}-\angle{x^0}.
\end{equation}

In the case of high SNR, where $M$ is not necessarily large, we get the same result as in (\ref{eq:cwcderiv:phase_first_approx}), since for high enough SNR the elements of signal-noise ($\psi^1_{xv}$) and noise-noise ($\psi^1_{vv}$) multiplications are neglectable relative to the signal-signal elements ($\psi^1_{xx}$).

Substituting (\ref{eq:cwcderiv:input}) and (\ref{eq:cwcderiv:phase_first_approx}) back into (\ref{eq:usage:optcomp:ybar_iter}), we get the result of the first aggregation step
\begin{equation}\label{eq:cwcderiv:vec_first}
    \hat{\bar{y}}^1_m=\hat{\bar{x}}^1g_{\tau_{m,0}}\eta_m + \hat{\bar{v}}_m^1,
\end{equation}
where
\begin{gather*}
    \hat{\bar{x}}^1=x^0+x^1e^{-j\Delta\phi^1}\approx \left(\left|x^0\right|+\left|x^1\right|\right)e^{j\angle x^0},\\
    \hat{\bar{v}}_m^1=v_m^0+v_m^1e^{-j\Delta\phi^1}.
\end{gather*}

In (\ref{eq:cwcderiv:vec_first}) we see that after the first aggregation step of CWC, the result is of the same form as the input (\ref{eq:cwcderiv:input}) and that the phase of the aggregated transmit signal equals that of the first window, i.e., $\angle{\hat{\bar{x}}^1}=\angle{x^0}$. Since the following steps of CWC are identical, it is easy to extrapolate that the result of each step will maintain the same form. Therefore, the phase-difference at the $d$th step will be given by
\begin{equation*} \label{eq:cwcderiv:phase_aprox}
    \Delta\hat{\phi}^d\approx \angle{x^d} - \angle{\hat{\bar{x}}^{d-1}}\approx\angle{x^d}-\angle{x^0}.
\end{equation*}






\bibliographystyle{IEEEtran}

\bibliography{bibtex/bib/IEEEabrv,bibtex/bib/IEEEexample,bibtex/bib/mybibliography}

\newpage

\end{document}